\documentclass[12pt]{iopart}
\usepackage{graphicx}
\usepackage{bm}
\begin{document}

\title[]
{Tuning the electrical resistivity of semiconductor thin films
by nanoscale corrugation}

\author{Shota Ono and Hiroyuki Shima}

\address{Department of Applied Physics, Graduate School of Engineering, Hokkaido University, Sapporo. 060-8628 Japan}
\ead{shota-o@eng.hokudai.ac.jp}
\ead{shima@eng.hokudai.ac.jp}
\begin{abstract}
The low-temperature electrical resistivity of corrugated semiconductor films is theoretically
considered. Nanoscale corrugation enhances the electron-electron scattering contribution to
the resistivity, resulting in a stepwise resistivity development with increasing corrugation
amplitude. The enhanced electron scattering is attributed to the curvature-induced
potential energy that affects the motion of electrons confined to a thin curved film.
Geometric conditions and microscopic mechanism of the stepwise resistivity are
discussed in detail.
\end{abstract}

\maketitle

\section{Introduction}
Nanostructures with curved geometry have attracted broad interest in the last decade.
Successful fabrication of corrugated semiconductor films\cite{prinz1,prinz2}, M\"{o}bius NbSe$_3$ stripes\cite{tanda},
peanut-shaped C$_{60}$ polymers\cite{onoe1,onoe2,onoe3}, and other exotic nanomaterials
with complex geometry\cite{exp4,exp5,exp6,gao,motojima,wang,cgaas,exp7,suspended}
has triggered the development of next-generation nanodevices.
Moreover, nanostructures provide an experimental platform for exploring the effects of surface curvature on the nature
of conducting electrons confined to low-dimensional systems. An important consequence of non-zero surface curvature
is the occurrence of a curvature-induced effective potential. It was theoretically suggested\cite{Jensen,costa,Kaplan,Jaffe}
that an electron moving in a thin curved layer experiences potential energy whose sign and magnitude depend on the local geometric curvature.
Such a curvature-induced potential has been observed to cause many intriguing phenomena\cite{ex1,ex2,ex3,ex4,
ex5,ex6,ex7,ex8,ex9,ex10,ex11,ex12,ex13,ex14,ex15,ex16,ex17,ex18,ex19,ex20}:
such as bound states of non-interacting electrons in deformed cylinders\cite{cant,marchi,taira}
and energy band gaps in periodic curved surfaces\cite{aoki,fujita,koshino,fujita2}.
Quite recently, surface curvature was found to markedly affect interacting electrons in the quasi-one dimension,
resulting in a significant shift in the Tomonaga-Luttinger exponent of
thin hollow cylinders subject to periodic surface deformation\cite{shima}.

In the present study, we demonstrate an alternative consequence of surface curvature,
which manifests in interacting electron systems. We consider the low-temperature resistivity of two-dimensional
corrugated semiconductor films and show that nanoscale corrugation considerably enhances the resistivity of the films.
This resistivity enhancement is attributed to contributions of electron-electron Umklapp scattering processes.
When the amplitude of corrugation takes specific values determined by the Fermi energy,
Umklapp processes due to the curvature-induced periodic potential cause a change in the total electron momentum,
resulting in a significant increase in the resistivity.
The period and amplitude of corrugations associated with the enhanced resistivity are within the realm of existing experiments\cite{Messica},
confirming the relevance of our theoretical predictions to curved-structure-based application technology. 

This paper is organized as follows. In section 2, we provide an outline of the derivation of the Sch\"{o}dinger equation for a curved surface
on the basis of the da Costa approach\cite{costa}. In section 3 and section 4, we summarize the 
formula for calculating the resistivity that is affected by electron-electron scattering
on the periodically corrugated surface at low temperature.
In section 5 and section 6, we present the numerical results and discussions, respectively. Finally, in section 7, we conclude the paper. 
\begin{figure}
\center
\includegraphics[scale=0.7,clip]{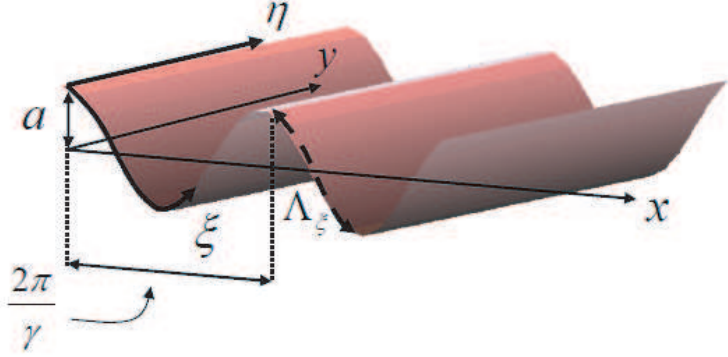}
\caption{\label{fig:curve}  
Schematic illustration of thin corrugated surface represented by $z=a\cos(\gamma x)$.
Here, $a$ is the corrugation amplitude and ${2\pi}/{\gamma}$ is the period of corrugation. 
The curvilinear coordinates ($\xi$,$\eta$) and the period of a curvature-induced potential 
$U(\xi)$, denoted by $\Lambda_\xi$, are indicated. See text for the definition of $U(\xi)$.}
\end{figure}
\section{Electron eigenstates in nanocorrugated films}
\begin{figure}
\center
\includegraphics[scale=1.0,clip]{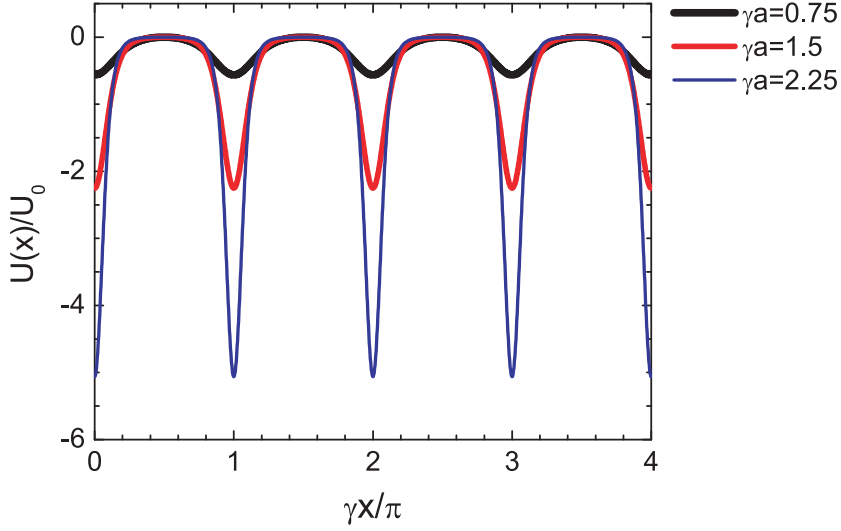}
\caption{\label{fig:pot} 
Spatial profile of $U(x)$ in units of $U_0\equiv{\hbar^2 \gamma^2}/{8m^*}$.
Downward peaks of $U(x)$ at ${\gamma x}/{\pi}=\ell\ (\ell=0, 1, 2, \cdots)$ grow with
increasing $a$.}
\end{figure}

In this section, we outline the mathematical description of electrons confined to a periodically curved surface.
We assume a thin conducting layer to which the motion of an electron is confined.
The layer is corrugated in the $x$-direction with period $2\pi/\gamma$ but remains flat in the $y$-direction.
The height of the layer is expressed as 
\begin{equation}
 z=a\cos(\gamma x),
\end{equation}
where $a$ is the amplitude of corrugation (see Fig.1).
We assume that the thickness of the layer is spatially uniform and sufficiently small
to increase excitation energies in the normal direction far beyond those in the tangential direction.
Under these conditions, we obtain the Sch\"{o}dinger equation for electrons propagating in the corrugated layer as\cite{costa}
\begin{eqnarray}
 -\frac{\hbar^2}{2m^*}\left[\frac{1}{w(x)}\frac{\partial}{\partial x}\left(\frac{1}{w(x)}\frac{\partial}{\partial x}\right)
 +\frac{\partial^2}{\partial y^2}\right]\psi(x,y)+U(x)\psi(x,y) = E\psi(x,y), \label{eq:xy} \nonumber\\
\end{eqnarray}
where $m^*$ is the effective electron mass and $w(x)=\sqrt{1+(a\gamma \sin(\gamma x))^2}$.
The salient feature of Eq.(\ref{eq:xy}) is the presence of an attractive potential $U(x)$ defined by 
\begin{equation}
   U(x)=-\frac{\hbar^2}{8m^*}\frac{[a\gamma^2\cos(\gamma x)]^2}{\left[1+(a\gamma\sin(\gamma x))^2\right]^{3}}\label{eq:Ux},
\end{equation}
which results from the non-zero surface curvature of the system.
(In fact, $U(x)\equiv 0$ if $a\equiv0$, i.e. a flat surface.)
Spatial profiles of $U(x)$ in units of $U_0={\hbar^2\gamma^2}/{8m^*}$ are shown in Fig.\ref{fig:pot},
where several values of $a$ are selected. Downward peaks are formed at $x=\pm \ell \pi/\gamma\ (\ell=0,1,2,\cdots)$,
where the layer height is either maximum ($z=+a$) or minimum ($z=-a$).
It is noteworthy that the $x$-dependence of $U(x)$ deviates considerably from a sinusoidal curve,
whereas the surface corrugation is exactly sinusoidal. 

Equation (\ref{eq:xy}) is simplified by using new variables (see Fig.\ref{fig:curve}): 
\begin{eqnarray}
 \xi =\int_{0}^{x}w(x')dx',\ \eta = y. \label{eq:xi}
\end{eqnarray}
Substituting them into Eq.(\ref{eq:xy}) yields an alternative form of the Schr\"{o}dinger equation
\begin{eqnarray}
 -\frac{\hbar^2}{2m^*}\left(\frac{\partial^2}{\partial \xi^2}+\frac{\partial^2}{\partial \eta^2} \right)\psi(\xi,\eta)
 +U(\xi)\psi(\xi,\eta) = E\psi(\xi,\eta), \label{eq:xieta} 
\end{eqnarray}
which has the solution of the form $\psi(\xi,\eta)=X(\xi)Y(\eta)$,
with $X(\xi)$ and $Y(\eta)$ satisfying the equations
\begin{eqnarray}
 -\frac{\hbar^2}{2m^*} \frac{\partial^2}{\partial \xi^2}X(\xi)+U(\xi)X(\xi) = E_\xi X(\xi),   \label{eq:Schrodinger}\\
 -\frac{\hbar^2}{2m^*}\frac{\partial^2}{\partial \eta^2}Y(\eta)=E_\eta Y(\eta) \label{eq:sc_eta},
\end{eqnarray}
where $E_\xi +E_\eta =E$. From Eq.(\ref{eq:sc_eta}), we obtain
$Y(\eta)\propto e^{ik^\eta\eta}$.
Equation (\ref{eq:Schrodinger}) is solved by using the Fourier series expansions
\begin{equation}
 X(\xi)=\sum_{k_\xi} \alpha_{k^\xi} e^{ik^\xi \xi},\ k^\xi=\frac{2\pi}{L_\xi}n_\xi, \label{eq:wavefuncX}
\end{equation}
and 
\begin{eqnarray}
 U(\xi)= \sum_G U_{G} e^{iG \xi},\ G=\frac{2\pi}{\Lambda_\xi}n, \label{eq:potentialU}
\end{eqnarray}
where $n_\xi, n$ are integers; ${L_\xi=\int_{0}^{L_x}w(x)dx}$ and $L_x$ represent the length of the layer along
the $\xi$- and $x$-coordinate axes, respectively, and $\Lambda_\xi=\int_{0}^{\pi/\gamma}w(x)dx$
equals one period of $U(\xi)$ (see Fig.\ref{fig:curve}). Substituting Eqs.(\ref{eq:wavefuncX}) and (\ref{eq:potentialU})
into Eq.(\ref{eq:Schrodinger}), we obtain the secular equation 
\begin{equation}
 \biggr(\epsilon_{k^\xi-G}^0-E_\xi \biggr)\alpha_{k^\xi-G} +
 \sum_{G'}U_{G'-G}\alpha_{k^\xi-G'}=0 \ \ \rm{for\ every}\ {\it k^\xi}, \label{eq:diag}
\end{equation}
where $\epsilon_{k^\xi-G}^0=\frac{\hbar^2}{2m^*}(k^\xi-G)^2$.
Numerical diagonalization of Eq.(\ref{eq:diag}) yields $E_\xi$ and $X(\xi)$ that satisfy Eq.(\ref{eq:Schrodinger}),
which completes calculations of eigenstates of electrons confined to a thin corrugated layer. 
\begin{figure}
\center
\includegraphics[scale=0.9,clip]{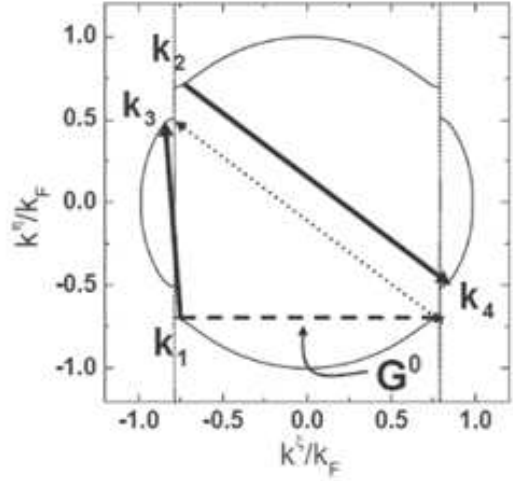}
\caption{\label{fig:fermi0}
Schematic illustration of first-order Umklapp scattering process from between
two-electron states $(\bm{k}_1,\bm{k}_2)$ and $(\bm{k}_3,\bm{k}_4)$. Gaps open at $k^\xi/k_F= \pm 0.79$,
whose positions are determined by $E_F$ and $a$.
Here, as an example, we set $E_F/U_0=3.13$ and $k_F a=1.32$.} 
\end{figure}
\section{Two-electron scattering processes: Umklapp and normal processes}
The objective here is to determine the effect of electron-electron scattering processes on the low-temperature
resistivity of thin corrugated layers. We assume a two-electron scattering process that transforms
a pair of electron states ($\bm{k}_1,\bm{k}_2$) into ($\bm{k}_3,\bm{k}_4$), where $\bm{k}_i=(k_i^\xi,k_i^\eta)$
is the wave vector of the $i$th degenerate
eigenstate that belongs to a given Fermi energy $E_F$. For the law of momentum conservation to hold, conditions 
\begin{equation}
 k_3^\xi+k_4^\xi=k_1^\xi+k_2^\xi + \frac{2\pi}{\Lambda_\xi}m, \label{eq:kxi}
\end{equation}
with arbitrary integer $m$, and 
\begin{equation}
 k_3^\eta+k_4^\eta=k_1^\eta+k_2^\eta. \label{eq:keta}
\end{equation}
must be satisfied. The last term on the right side of Eq.(\ref{eq:kxi}) is attributed to the periodic structure
of $U(\xi)$ that yields a {\it reciprocal lattice vector} $G^0\equiv {2\pi}/{\Lambda_\xi}$.
A schematic illustration of a two-electron scattering process is shown in Fig.\ref{fig:fermi0}.
Thin curves on the $k^\xi$-$k^\eta$ plane represent the Fermi surface
for $E_F/U_0$ = 3.13 and $k_F a$ = 1.32; these thin curves show gaps at $k^\xi \simeq \pm 0.79 k_F$,
which are caused by the periodic potential $U(\xi)$. Here, we assume that in the absence of corrugation
(i.e. $a\equiv 0$), the Fermi surface shows an exact circle on the $k^\xi$-$k^\eta$ plane.
As shown in Fig.\ref{fig:fermi0}, two eigenstates located at $\bm{k}_1$ and $\bm{k}_2$,
in the vicinity of a corrugation-induced gap, are transformed into $\bm{k}_3$ and $\bm{k}_4$, respectively.
It should be noted that both relations (\ref{eq:kxi}) and (\ref{eq:keta}) hold for the process shown in Fig.\ref{fig:fermi0},
where the integer $m$ in Eq.(\ref{eq:kxi}) takes the value of $m=1$. Hereafter,
a two-electron scattering process involving an integer $m\ne 0$ is termed the $m$th Umklapp process;
if $m=0$, we refer to it as a normal scattering process. 
\section{Boltzmann transport equation}
Contributions of two-electron scattering to the resistivity at low temperature $T$
are given by the Boltzmann transport equation\cite{ziman}. We assume that $k_BT\ll E_F$ and the Fermi surface has a circular shape on the 
$k^\xi$-$k^\eta$ plane if $a\equiv 0$. Then, we can prove that\cite{uryu}
\begin{eqnarray}
 \rho(T)&=&\rho_{c}\biggr(\frac{k_BT}{E_F}\biggr)^2 \frac{h}{e^2}, \\
 \rho_{c}&=&\frac{1}{192\pi^2\hbar^2} \sum_{m=-\infty}^\infty \int d\bm{q}
  \frac{\big(\vert M_a \vert^2+\frac{1}{2}\vert M_a-M_b \vert^2\big)(\Delta \bm{v}\cdot \bm{u})^2} 
  {|\bm{v}_1\times \bm{v}_3||\bm{v}_2\times \bm{v}_4|},  \label{eq:rho_c}
\end{eqnarray}
where $k_B$ is the Boltzmann constant, $\bm{v}_i$ is the group
velocity of the electron belonging to the eigenstate $\bm{k}_i$,
$\Delta \bm{v}=(\bm {v}_3+\bm {v}_4)-(\bm{v}_1+\bm {v}_2)$,
and $\bm {u}$ is the applied electric field.
Integration over $\bm{q}\equiv \bm{k}_3-\bm{k}_1$ in Eq.(\ref{eq:rho_c}) is carried out for all possible
$\bm{k}_1$ and $\bm{k}_3$ that satisfy relations (\ref{eq:kxi}) and (\ref{eq:keta}) for a fixed $m$.
The transition probability $M_j\ (j=a,b)$ is given by
\begin{eqnarray}
M_j&=&\sum_{G_1}\cdots \sum_{G_4}\tilde{u}(\bm{K}_j)\alpha_{k_3^\xi-G_3}^* \alpha_{k_4^\xi-G_4}^* 
\alpha_{k_2^\xi-G_2}\alpha_{k_1^\xi-G_1}\ (j=a,b)  \label{eq:M'} \\  
 \bm{K}_a&=&\bm{k}_3-\bm{k}_1-\bm{G}_3+\bm{G}_1, \nonumber\\
 \bm{K}_b&=&\bm{k}_4-\bm{k}_1-\bm{G}_4+\bm{G}_1,\nonumber
\end{eqnarray}
where $\tilde{u}(\bm{K})$ is the Fourier transform of the screened Coulomb potential\cite{uryu}
and $\bm{G}_i=(n_iG^0,0)$ in terms of the ($\xi,\eta$) coordinates. As is clear from Eq.(\ref{eq:M'}),
summations with respect to $G_i\equiv n_iG^0$ are carried out for all $n_i$ values under the constraints 
\begin{equation}
 n_3+n_4=n_1+n_2+ m,
\end{equation}
where $m$ is fixed by the summation index in Eq.(\ref{eq:rho_c}).
In the actual calculation, we employed material constants of GaAs/Al$_x$Ga$_{1-x}$As heterostructures:
$E_F=10$ meV, $m^*=0.067m_0$, and $\epsilon=13.2\epsilon_0$ with a bare electron mass $m_0$ and
the dielectrical constant of vaccum $\epsilon_0$. In such heterostructures,
the Fermi energy exists near the $\Gamma$ point, which justifies our assumption of an isotropic Fermi surface. 
\begin{figure}
\center
\includegraphics[scale=1.1,clip]{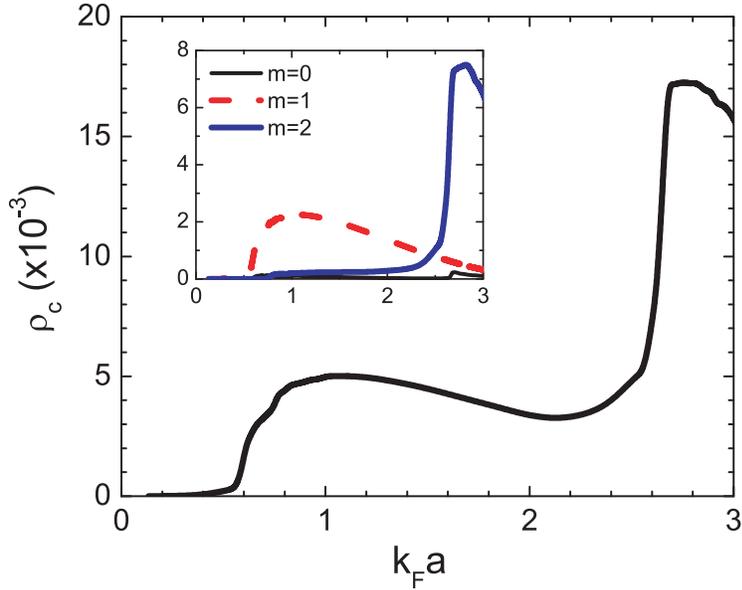}
\caption{\label{fig:rho_c1} Stepwise increase in $\rho_c$ with corrugation amplitude $a$.
$\rho_c$ shows sudden jumps at $k_F a$ = 0.6 and 2.6.
Inset: contribution to $\rho_c$ from the normal ($m=0$) process and Umklapp process of the
first ($m=1$) and second ($m=2$) orders.}
\end{figure}
\section{RESULTS}
Figure \ref{fig:rho_c1} shows the $a$-dependence of $\rho_c$ at $\gamma=0.15$ nm$^{-1}$.
The surprising observation in Fig.\ref{fig:rho_c1} is a stepwise increase in $\rho_c$ with $a$,
in which $\rho_c$ exhibits sudden jumps at specific values of $a$ given by $k_F a$ = 0.6 and 2.6.
The non-monotonic behaviour of $\rho_c$ plotted in Fig.\ref{fig:rho_c1} suggests the possibility of tuning the resistivity of actual
low-dimensional nanostructures by imposing surface corrugation.
In fact, our results suggest that the corrugation-induced increase in the resistivity
observed in $\rm GaAs$-based nanocorrugated films with $a=$ 5 nm and $2\pi/\gamma=$ 40 nm is of the order of $ \rho_c \sim 3\times 10^{-3}$,
which is within the range obtained by the measurement technique used in Refs.\cite{Messica}. 
\begin{figure}
\center
\includegraphics[scale=1.1,clip]{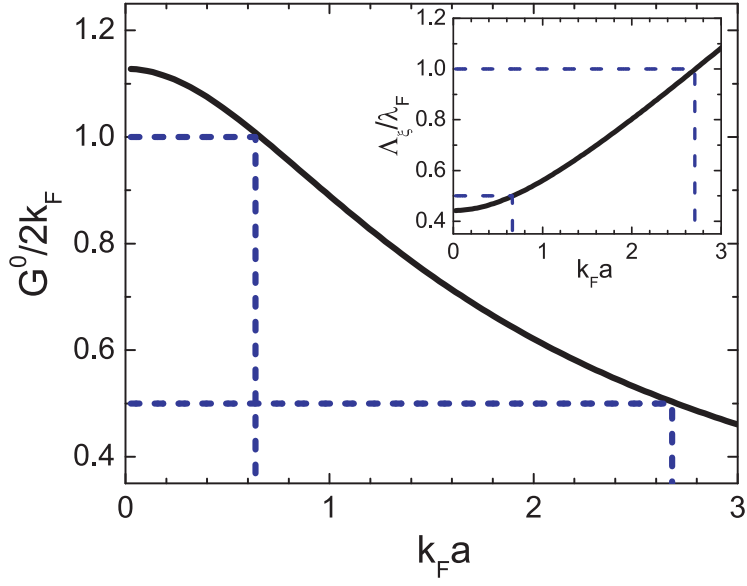}
\caption{\label{fig:gkf} 
Monotonic decreasing behaviour of $G^0$ with increasing $a$.
$G^0/{2k_F}$ takes values of $1$ and $1/2$ at $k_F a=$ 0.6 and 2.6, respectively.
Inset: $a$-dependence of $\Lambda_\xi$ in units of Fermi wavelength $\lambda_F$.}
\end{figure}

To determine the physical origin of jumps, we decompose the summation shown in Eq.(\ref{eq:rho_c}) with respect to $m$
and separately plot three dominant components consisting of $\rho_c$, as shown in the inset
of Fig.\ref{fig:rho_c1}: the thin solid line shows the contribution of the normal scattering process ($m=0$),
the thin dotted line shows the contribution of the first-order Umklapp process ($m=1$),
and the thick solid line shows the contribution of the second-order Umklapp process ($m=2$).
The plot in the inset shows that the significant increase in the Umklapp contributions results in the jump in $\rho_c$.
The mechanism of the increase in Umklapp contributions is explained in detail in the next section. 

We also find that specific values of $k_F a$ which cause the jump in $\rho_c$ correspond to those of
$k_F a$ which satisfy the relation $G^0/{2k_F}=1/p \ (p=1,2,\cdots)$, or equivalently,
$\Lambda_\xi=p\lambda_F/2$, where $\lambda_F$ is the Fermi wavelength.
Figure \ref{fig:gkf} shows the $a$-dependence of $G^0/{2k_F}$; it decreases monotonically with $k_F a$,
because $G^0\propto \Lambda_\xi^{-1}$ and $\Lambda_\xi$ increases with $a$ (see Fig.\ref{fig:gkf}).
We see from Fig.\ref{fig:gkf} that $G^0/{2k_F}$ takes
the values of $1$ and $1/2$ at $k_F a=$ 0.6 and $2.6$, respectively,
which cause the jumps in $\rho_c$ as shown in Fig.\ref{fig:rho_c1}. It should be noted that at these values of $k_F a$,
the radius of the Fermi circle on the $k^\xi$-$k^\eta$ plane becomes equal to $G^0$ or $G^0/2$.
As a result, the gaps open at $k^\xi=\pm k_F$ (as well as $\pm k_F/2$), i.e. at both ends of the Fermi circle.
These gaps lead to the increase in the Umklapp contribution to $\rho_c$, as elucidated in the next section. 
\section{DISCUSSIONS}
\label{dis}
The jumps in $\rho_c$ shown in Fig.\ref{fig:rho_c1} result from
the following three conditions: i) enhanced transition probabilities $M_j$ in the integrand of $\rho_c$,
ii) divergence of the density of states $(|\bm{v}_1\times \bm{v}_3||\bm{v}_2\times \bm{v}_4|)^{-1}$ (see Eq.(\ref{eq:rho_c})),
and iii) occurrence of gaps at both the ends of the Fermi circle (i.e. at $k^\xi=\pm k_F$).
To present a concise argument, we consider first-order Umklapp contributions,
i.e. components related to $m=1$ in expression (\ref{eq:rho_c}).
(An analogous discussion for the case of $m=2$ is available.) 
\begin{figure}
\center
\includegraphics[scale=1.1,clip]{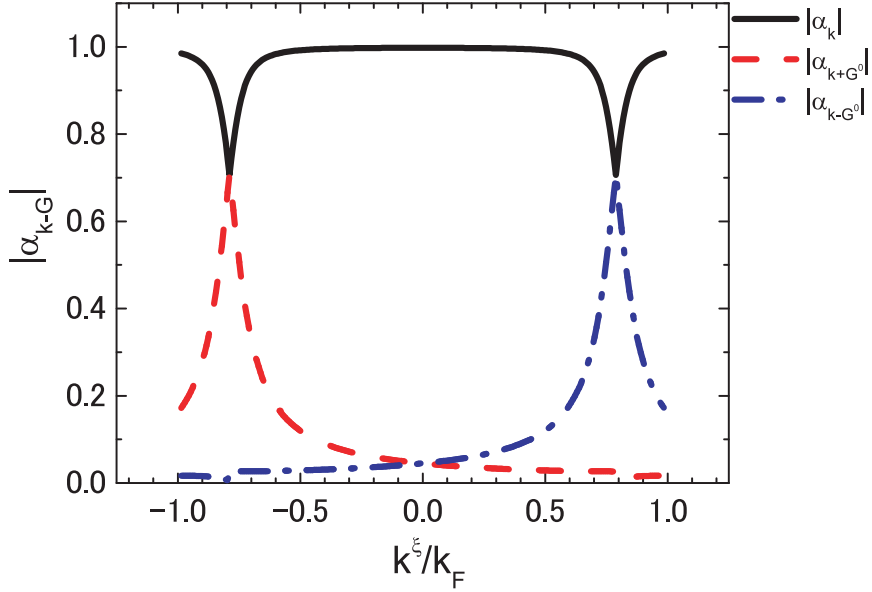}
\caption{\label{fig:wavek} 
Profile of expansion coefficients $\vert\alpha_{k^\xi}\vert$ as a function of $k^\xi$;
see Eq.(\ref{eq:wavefuncX}) for its definition. The magnitude of $\vert\alpha_{k^\xi\pm G^0}\vert$
is comparable to that of $\vert\alpha_{k^\xi}\vert$ near $k^\xi/k_F=\mp0.79$, 
respectively, at which gaps open as shown Fig.\ref{fig:fermi0}.}
\end{figure}
We know that $M_j$ includes the term $\propto\alpha_{k_3^\xi}^* \alpha_{k_4^\xi}^* \alpha_{k_2^\xi}\alpha_{k_1^\xi+G^0}$, for instance.
Here, $\vert\alpha_{k_1^\xi+G^0}\vert$ has a large value within a limited region $\Delta k_1^\xi$
centred at $k_1^\xi=-G^0/2$ (see Fig.\ref{fig:wavek}).
It is noteworthy that larger $\vert\alpha_{k_1^\xi+ G^0}\vert$ results in larger $M_j$; furthermore,
this scenario holds for other terms involving $\alpha_{k_i^\xi+G^0}\ (i=2,3,4)$.
Therefore, $M_j$ is enhanced when at least one of the four states,
$k_i^\xi$, is located with the corresponding region $\Delta k_i^\xi$ within which $\vert\alpha_{k_i^\xi+ G^0}\vert$ is large. 

\begin{figure}
\center
\includegraphics[scale=0.6,clip]{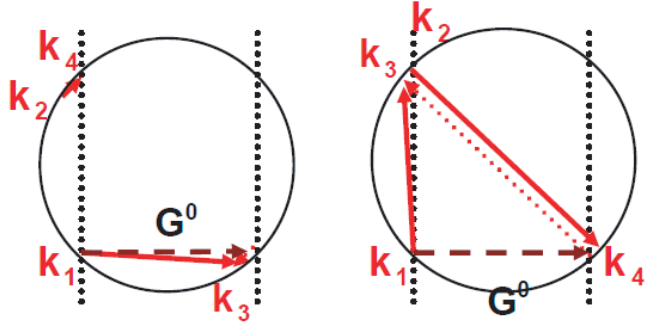}
\caption{\label{fig:scatt} 
Examples of Umklapp scattering process in which states $\bm{k}_2$ and $\bm{k}_4$ undergo
forward scattering (left) or backward scattering (right). All four states $\bm{k}_i\ (i=1,2,3,4)$ are constrained
to be located near the corrugation-induced gap denoted by the vertical dotted lines.
}
\end{figure}

Next, we consider the condition for the density of states $(|\bm{v}_1\times \bm{v}_3||\bm{v}_2\times \bm{v}_4|)^{-1}$
to diverge. From Ref.\cite{uryu}, it follows that 
\begin{eqnarray}
 \int \frac{d\bm{q}}{|\bm{v}_1\times \bm{v}_3||\bm{v}_2\times \bm{v}_4|}
  \propto \int \frac{d\theta_1d\theta_3}{\vert \sin \theta_{24}\vert}, \label{eq:dq_int}
\end{eqnarray}
where $\theta_1$ and $\theta_3$ are the polar angles of $\bm{k}_1$ and $\bm{k}_3$,
respectively, on the $k^\xi$-$k^\eta$ plane and $\theta_{24}$ is the relative angle
between $\bm{k}_2$ and $\bm{k}_4$. Expression (\ref{eq:dq_int}) implies that $\rho_c$ diverges when $\theta_{24}\simeq 0$ and $\theta_{24}\simeq \pi$
which correspond to forward and backward scattering, respectively, between the states of $\bm{k}_2$ and $\bm{k}_4$.
As a consequence, sudden jumps in $\rho_c$ are attributed to the Umklapp process that involves forward and backward scattering.
Figure \ref{fig:scatt} shows some examples of relevant scattering processes that satisfy conditions i) and ii) mentioned earlier. 

\begin{figure}
\center
\includegraphics[scale=0.6,clip]{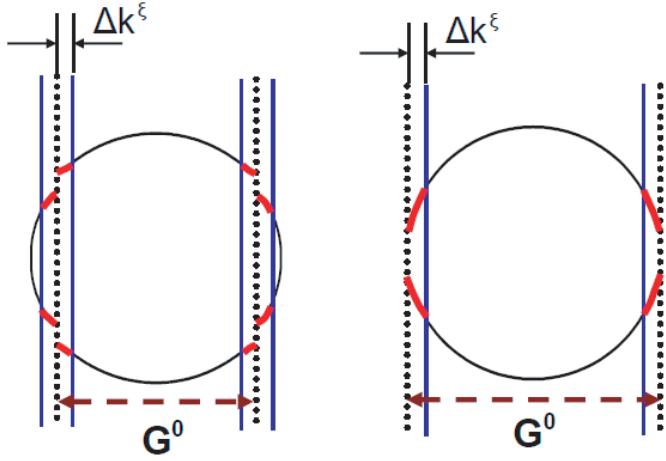}
\caption{\label{fig:gap}
Fermi circles showing gaps at (a) $k^\xi=\pm 0.79k_F$ and (b) $k^\xi=\pm k_F$.
The portion of the thick arcs (coloured in red) indicates eigenstates lying interior to the region of $\Delta k^\xi$,
i.e. states in the vicinity of gaps, which significantly contribute to $\rho_c$.}
\end{figure}

Finally, we comment on condition iii), that is, gaps in the Fermi surface should be positioned at both the ends of the Fermi circle,
$k^\xi=\pm k_F$. Figure \ref{fig:gap} shows two Fermi circles, in each of which gaps open at (a) $k^\xi=\pm 0.79k_F$ and (b) $k^\xi=\pm k_F$.
Circular thick arcs (coloured in red) indicate positions of eigenstates interior to the region of $\Delta k^\xi$,
i.e. states in the vicinity of gaps. From earlier discussions, it follows that Umklapp scattering processes
involving the states indicated by thick curves significantly contribute to $\rho_c$.
It should be noted that such relevant processes are achieved more frequently in the Fermi circle
shown in Fig.\ref{fig:gap}(b) than in that shown in Fig.\ref{fig:gap}(a),
since the length of the thick curve in the former is larger than that in the latter.
This proves that condition iii) is essential for the occurrence of the sudden jump in $\rho_c$. 
\section{CONCLUSION}
In conclusion, we have shown that the electrical resistivity of nanocorrugated semiconductor films
exhibits a stepwise increase with the corrugation amplitude.
Corrugation amplitudes that lead to resistivity jumps are determined by the relation $G^0/2k_F=1/p\ (p=1,2,\cdots)$,
where $k_F$ is the Fermi wave vector and $G^0$ is the corrugation-induced reciprocal lattice vector
associated with the curvature-driven periodic potential $U$.
We have proved that the resistivity jumps originate from the increased contribution of an Umklapp scattering process.
The requisite corrugation amplitude and period for the resistivity jumps to be observable are within the realm of laboratory conditions,
which confirms that our theoretical prediction can be verified experimentally. 
\section*{Acknowledgments}
We would like to thank K. Yakubo, S. Nishino, H. Suzuura, and S. Uryu
for useful discussions and suggestions. This study was supported by a Grant-in-Aid for Scientific
Research from the MEXT, Japan. One of the authers (H.S.) is thankful for the financial support
from Executive Office of Research Strategy in Hokkaido University.
A part of numerical simulations were carried out using the facilities of the Supercomputer
Center, ISSP, University of Tokyo.
\appendix
\section{The Schr\"{o}dinger equation of corrugated nanofilms}
This appendix describes the derivation of Eq.(\ref{eq:xy}), i.e. the Schr\"{o}dinger equation
that describes the motion of electrons confined to a thin curved film.
We assume that an electron located on a general curved surface is parameterised by
$\bm{r}=\bm{r}(q_1,q_2)$, where $\bm{r}$ is the position vector of an arbitrary point on the surface.
According to the da Costa approach\cite{costa}, the Schr\"{o}dinger equation on the curved surface is given by 
\begin{eqnarray}
 -\frac{\hbar^2}{2m^*}\sum_{i,j=1}^2\frac{1}{\sqrt{g}} \frac{\partial}{\partial q_i}
 \left(\sqrt{g}g^{ij}\frac{\partial}{\partial q_j} \psi(q_1,q_2) \right)+U(q_1,q_2)\psi(q_1,q_2) \label{eq:chi_S}
 =E\psi(q_1,q_2), \nonumber\\
\end{eqnarray}
where $m^*$ is the effective mass of the electron, $g_{ij}$ is the metric tensor defined by
$g_{ij}=\frac{\partial \bm{r}}{\partial q_i}\cdot \frac{\partial \bm{r}}{\partial q_j} \ (i,j=1,2)$,
$g^{ij}$ is the inverse of $g_{ij}$, and $g=\det (g_{ij})$\cite{shima_naka}.
The second term on the left side of Eq.(\ref{eq:chi_S}) is the effective potential generated by the curvature of the system and is given by 
\begin{equation}
  U(q_1,q_2) = -\frac{\hbar^2}{2m^*}\left(\left[\frac{1}{2}\Tr (\alpha_{ij})\right]^2-\det(\alpha_{ij})\right)
  =-\frac{\hbar^2}{2m^*}({\cal H}^2-{\cal K}) \label{eq:uu}
\end{equation}
where ${\cal H}$ and ${\cal K}$ are the so-called mean curvature and Gauss curvature of the surface, respectively.
$\alpha_{ij}$ is the Weingarten curvature matrix expressed as 
\begin{eqnarray}
\left(
\begin{array}{cc}
 \alpha_{11} & \alpha_{12} \\
 \alpha_{21} & \alpha_{22}
\end{array}
\right)
=\frac{1}{g}
\left(
\begin{array}{cc}
 g_{12}K_{12}-g_{22}K_{11} & g_{12}K_{11}-g_{11}K_{12} \\
 g_{12}K_{22}-g_{22}K_{12} & g_{12}K_{12}-g_{11}K_{22}  \label{eq:aij}
\end{array}
\right),
\end{eqnarray}
in which $K_{ij}$ are the coefficients of the second fundamental form, $K_{ij}=\bm{n}\cdot \frac{\partial^2\bm{r}}{\partial q_i\partial q_j}$,
where $\bm{n}$ is the unit vector normal to the surface, 
\begin{eqnarray}
 \bm{n}={\frac{\partial \bm{r}}{\partial q_1} \times \frac{\partial \bm{r}}{\partial q_1}}\biggr/
 {\left\vert \frac{\partial \bm{r}}{\partial q_1} \times \frac{\partial \bm{r}}{\partial q_1} \right\vert}.
\end{eqnarray}

Next, we consider a periodically corrugated surface parameterised by 
\begin{equation}
 \bm{r}=x\bm{e}_x+y\bm{e}_y+f(x)\bm{e}_z, \label{eq:ff}
\end{equation}
where $f(x)$ is a periodic function. From the definition of $g_{ij}$, we obtain 
\begin{eqnarray}
g_{ij}=
\left(
\begin{array}{cc}
1+f_x^2 & 0 \\
0 & 1
\end{array}
\right),\ \ 
g^{ij}=
\left(
\begin{array}{cc}
\frac{1}{1+f_x^2} & 0 \\
0 & 1
\end{array}
\right).
\end{eqnarray}
The unit normal $\bm{n}$ is expressed as 
\begin{eqnarray}
 \bm{n}=\left( \frac{-f_x}{\sqrt{1+f_x^2}},0,\frac{1}{\sqrt{1+f_x^2}}\right), \label{eq:nn}
\end{eqnarray}
which yields 
\begin{eqnarray}
K_{ij}=
\frac{1}{(1+f_x^2)^{3/2}}
\left(
\begin{array}{cc}
f_{xx} & 0 \\
0 & 0
\end{array}
\right)  
{\rm and\ thus}\
\alpha_{ij}=
\frac{1}{(1+f_x^2)^{3/2}}
\left(
\begin{array}{cc}
-f_{xx} & 0 \\
0 & 0
\end{array}
\right).
\nonumber\\
\end{eqnarray}
Consequently, by substituting the above results into Eq.(\ref{eq:uu}) and then assuming $f(x)=a\cos (\gamma x)$,
we obtain the Schr\"{o}dinger equation, i.e. Eq.(\ref{eq:xy}).
It should be emphasized that though our study focuses on unidirectionally
corrugated films, our theoretical approach is applicable to a wide variety
of nanostructures with periodically curved geometry, such as those suggested in Refs\cite{MCNT,MCNTshima}. 


\section*{References}
\begin{thebibliography}{99}
\bibitem{prinz1}Prinz V Ya, Gr\"{u}tzmacher D, Beyer A, David C, Ketterer B
and Deckardt E 2001 {\it Nanotechnology} {\bf 12} 399
\bibitem{prinz2} Prinz V Ya 2006 {\it phys. stat. sol.} (b) {\bf 243} 3333
\bibitem{tanda} Tanda S, Tsuneta T, Okajima Y, Inagaki K, Yamaya K and Hatakenaka N 2002 {\it Nature} (London) {\bf 417} 397 
\bibitem{onoe1} Onoe J, Nakayama T, Aono M and Hara T 2003 {\it Appl. Phys. Lett.} {\bf 82} 595
\bibitem{onoe2} Onoe J, Ito T, Kimura S, Ohno K, Noguchi Y and Ueda S 2007 {\it Phys. Rev.} B {\bf 75} 233410
\bibitem{onoe3} Onoe J, Ito T and Kimura S 2008 {\it J. Appl. Phys.} {\bf 104} 103706
\bibitem{exp4} Lorke A, Bohm S and Wegscheider W 2003 {\it Superlattices Microstruct.} {\bf 33} 347
\bibitem{exp5} McIlroy D N, Alkhateeb A, Zhang D, Aston D E,
Marcy A C and Norton M G 2004 {\it J. Phys.: Condens. Matter} {\bf 16} R415
\bibitem{exp6} Sano M, Kamino A, Okamura J and Shinkai S 2004 
{\it Science} {\bf 293} 1299
\bibitem{gao} Gao P X, Ding Y, Mai W J, Hughes W L, Lao C S and
Wang Z L 2005 {\it Science} {\bf 309} 1700
\bibitem{motojima} Yang S, Chen X, Motojima S and Ichihara M 2005 {\it Carbon} {\bf 43} 827
\bibitem{wang} Wang L, Major D, Paga P, Zhang D, Norton M G and McIlroy D N 2006 {\it Nanotechnology} 17 S298
\bibitem{cgaas} Gong Z, Niu Z and Fang Z 2006 {\it Nanotechnology} {\bf 17} 1140
\bibitem{exp7} Fujita T, Qian L H, Inoke K, Erlebacher J and Chen M W 2008 {\it Appl. Phys. Lett.} {\bf 92} 251902
\bibitem{suspended} Sainiemi L, Grigoras K and Franssila S 2009 {\it Nanotechnology} {\bf 20} 075306
\bibitem{Jensen} Jensen H and Koppe H 1971 {\it Ann. Phys.} {\bf 63} 586
\bibitem{costa} da Costa R C T 1981 {\it Phys. Rev.} A {\bf 23} 1982
\bibitem{Kaplan} Kaplan L, Maitra N T and Heller E J 1997 {\it Phys. Rev.} A {\bf 56} 2592
\bibitem{Jaffe} Schuster P C and Jaffe R L 2003 {\it Ann. Phys.} {\bf 307} 132
\bibitem{ex1} Mostafazadeh A 1996 {\it Phys. Rev.} A {\bf 54} 1165
\bibitem{ex2} Entin M V and Magarill L I 2001 {\it Phys. Rev.} B {\bf 64} 085330
\bibitem{ex3} Entin M V and Magarill L I 2002 {\it Phys. Rev.} B {\bf 66} 205308
\bibitem{ex4} Encinosa M and Mott L 2003 {\it Phys. Rev.} A {\bf 68} 014102
\bibitem{ex5} Bulaev D V, Geyler V A and Margulis V A 2004 {\it Phys. Rev.} B {\bf 69} 195313
\bibitem{ex6} Chaplik A V and Blick R H 2004 {\it New J. Phys.} {\bf 6} 33
\bibitem{ex7} Entin M V and Magarill L I 2004 {\it Europhys. Lett.} {\bf 68} 853
\bibitem{ex8} Olendski O and Mikhailovska L 2005 {\it Phys. Rev.} B {\bf 72} 235314
\bibitem{ex9} Gravesen J and Willatzen M 2005 {\it Phys. Rev.} A {\bf 72} 032108
\bibitem{ex10} Encinosa M 2006 {\it Phys. Rev.} A {\bf 73} 012102
\bibitem{ex11} Zhang E, Zhang S and Wang Q 2007 {\it Phys. Rev.} B {\bf 75} 085308
\bibitem{ex12} Atanasova V and Dandoloff R 2007 {\it Phys. Lett.} A {\bf 371} 118
\bibitem{ex13} Balakrishnan R and Dandoloff R 2008 {\it Nonlinearity} {\bf 21} 1
\bibitem{ex14} Olendski O and Mikhailovska L 2008 {\it Phys. Rev.} B {\bf 77} 174405
\bibitem{ex15} Atanasova V and Dandoloff R 2008 {\it Phys. Lett.} A {\bf 372} 6141
\bibitem{ex16} Ferrari G and Cuoghi G 2008 {\it Phys. Rev. Lett.} {\bf 100} 230403
\bibitem{ex17} Ferrari G, Bertoni A, Goldoni G and Molinari E 2008 {\it Phys. Rev.} B {\bf 78} 115326
\bibitem{ex18} Atanasova V and Dandoloff R and Saxena A 2009 {\it Phys. Rev.} B {\bf 79} 033404
\bibitem{ex19} Cuoghi G, Ferrari G and Bertoni A 2009 {\it Phys. Rev.} B {\bf 79} 073410
\bibitem{ex20} Atanasova V and Dandoloff R 2009 {\it Phys. Lett.} A {\bf 373} 716
\bibitem{cant} Cantele G, Ninno D and Iadonisi G 2000 {\it Phys. Rev.} B {\bf 61} 13730
\bibitem{marchi} Marchi A, Reggiani S, Rudan M and Bertoni A 2005 {\it Phys. Rev.} B {\bf 72} 035403
\bibitem{taira} Taira H and Shima H 2007 {\it Surf. Sci.} {\bf 601} 5270
\bibitem{aoki} Aoki H, Koshino M, Takeda D and Morise H 2001 {\it Phys. Rev.} B {\bf 65} 035102
\bibitem{fujita} Fujita N 2004 {\it J. Phys. Soc. Jpn.} {\bf 73} 3115
\bibitem{koshino} Koshino M and Aoki H 2005 {\it Phys. Rev.} B {\bf 71} 073405
\bibitem{fujita2} Fujita N and Terasaki O 2005 {\it Phys. Rev.} B {\bf 72} 085459
\bibitem{shima} Shima H, Yoshioka H and Onoe J {\it arXiv:}0903.0798
\bibitem{Messica} Messica A, Soibel A, Meirav U, Stern A, Shtrikman H, Umansky V and Mahalu D 1997
{\it Phys. Rev. Lett} {\bf 78} 705 
\bibitem{ziman} Ziman J M 2001 {\it Electrons and Phonons} Oxford University Press, New York
\bibitem{uryu} Uryu S and Ando T 2001 {\it Phys. Rev.} B {\bf 64} 195334
\bibitem{shima_naka} Shima H and Nakayama T 2009 {\it Higher Mathematics for
Physics and Engineering} (Springer-Verlag)
\bibitem{MCNT} Arias I and Arroyo M 2008 {\it Phys. Rev. Lett.} {\bf 100} 230403
\bibitem{MCNTshima} Shima H and Sato M 2008 {\it Nanotechnology} {\bf 19} 495705

\end{thebibliography}
\end{document}